\documentclass[aps,prd,reprint,superscriptaddress,nofootinbib,floatfix]{revtex4-1}
\usepackage{hyperref}
\usepackage{amsmath}
\usepackage{lineno}
\usepackage{graphicx}
\usepackage{color}

\newcommand\refsec[1]{\S\ref{sec:#1}}
\newcommand\refeq[1]{Eq.~(\ref{eqn:#1})}

\newcommand{\reffig}[1]{Fig.~\ref{fig:#1}}

\newcommand{\aap}{A\&A}
\newcommand{\mnras}{MNRAS}

\newcommand{\jcap}{{\em JCAP }}

\begin{document}
\graphicspath{{./images/}}

\title{External priors for the next generation of CMB experiments.}
\author{Alessandro Manzotti}
\email{mailto:manzotti.alessandro@gmail.com}
\affiliation{Department of Astronomy \& Astrophysics, University of Chicago, Chicago IL 60637}
\affiliation{Kavli Institute for Cosmological Physics, Enrico Fermi Institute, University of Chicago, Chicago, IL 60637}
\author{Scott Dodelson}

\affiliation{Fermilab Center for Particle Astrophysics, Fermi National Accelerator Laboratory, Batavia, IL 60510-0500}
\affiliation{Department of Astronomy \& Astrophysics, University of Chicago, Chicago IL 60637}
\affiliation{Kavli Institute for Cosmological Physics, Enrico Fermi Institute, University of Chicago, Chicago, IL 60637}

\author{Youngsoo Park}
\affiliation{Department of Physics, University of Arizona, Tucson, AZ 85721, USA}

\date{\today}
\begin{abstract}
Planned cosmic microwave background (CMB) experiments will improve what we know about neutrino physics, inflation, and dark energy. 
The low level of noise, together with improved angular resolution, will increase the signal to noise of the CMB polarized data as well as the reconstructed lensing potential of large scale structure. Projected constraints on cosmological parameters are tight, but these can be improved even further with information from external experiments. Here, we examine quantitatively the extent to which external priors can lead to improvement in projected constraints from a CMB-Stage IV (S4) experiment on neutrino and dark energy properties.
We find that CMB S4 constraints on neutrino mass could be strongly enhanced by external constraints on the cold dark matter density $\Omega_{c}h^{2}$ and the Hubble constant $H_{0}$. If polarization on the largest scales ($\ell<50$) will not be measured, an external prior on the primordial amplitude $A_{s}$ or the optical depth $\tau$ will also be important. A CMB constraint on the number of relativistic degrees of freedom, $N_{\rm eff}$, will benefit from an external prior on the spectral index $n_{s}$ and the baryon energy density $\Omega_{b}h^{2}$. Finally, an external prior on $H_{0}$ will help constrain the dark energy equation of state ($w$).
\end{abstract}

\maketitle

\section{Introduction}\label{sec:intro}
Since their earliest incarnations, Cosmic Microwave Background (CMB) experiments have been crucial in furthering our understanding of the Universe. They will maintain their role in the future, thanks to the unprecedented low level of noise and high resolution expected in the planned Stage IV (S4) experiment~\cite{2013arXiv1309.5383A}.
The recent past is reassuring. For example, every new generation of satellite experiments improved the level of sensitivity by almost a factor of ten compared to its predecessor, from the first generation instrument COBE to WMAP all the way to the current state of the art represented by Planck \cite{fixsen:1996,BlueBook2005, bennett:2003}. 
These led us from the first detection of anisotropy in the CMB temperature with COBE to a cosmic variance limited measurement of several acoustic peaks with Planck. With the improved sensitivity we have extended our understanding of the Universe so that we now have solid evidence for a flat geometry and a well tested $\Lambda$CDM model with percent level constraints on its parameters.
Similar progress has characterized ground-based CMB experiments, where the first detection of polarization anisotropy \cite{2002ApJ...568...38H} has been followed by a series of precise measurements of the damping tail~\cite{2011ApJ...739...52D,2011ApJ...743...28K} and of the projected gravitational potential~\cite{2011PhRvL.107b1301D,2012ApJ...756..142V}.

The planned CMB-S4 experiment will potentially measure the E-mode polarization with cosmic variance limited precision together with an order of magnitude improvement in B-mode measurement and lensing reconstruction. As has happened in the past, this new sensitivity together with the progress in the measurement of other cosmological probes will improve our understanding of dark matter, inflation, dark energy, neutrinos, and other Beyond the Standard Model physics.

The cosmological dependence on neutrinos is two-fold: the number of relativistic species $N_{\rm eff}$ in the early phase of the Universe affects the damping tail of the CMB, and the sum of the neutrino masses $\sum m_\nu$ affects the late-time growth of cosmic structure.
The sensitivity to $N_{\rm eff}$ is due to its effects on the expansion rate $H(z)$ because relativistic species, like neutrinos, are the main drivers of the cosmic expansion in the early Universe. This rate can be powerfully tested using the CMB, by carefully comparing the sound horizon scale, obtained from the CMB peaks positions, and the Silk damping scale (see \cite{2013arXiv1309.5383A}, \cite{2013PhRvD..87h3008H} and references therein). 
The future experiments, with their high resolution, will probe deep into the damping tale of the CMB power spectrum and will be sensitive to small variations from the canonical, 3-active neutrino prediction of $N_{\rm eff}=3.046$. On the other hand, the total mass of neutrinos, has a modest effect on the CMB because, for the range of masses allowed by recent constraints ($\sum m_{\nu}<230$ meV from \cite{planck-collaboration:2014}), neutrinos are still relativistic at the last scattering surface. However, massive neutrinos alter the growth of the large scale structure responsible for lensing of the CMB photons. Different neutrino masses consequently lead to different CMB lensing spectra. CMB-S4 will measure small scale temperature and polarization anisotropies with low noise, improving by at least one order of magnitude the current lensing reconstruction. This will turn into a precise constraints of the sum of neutrino masses with a possible hint of the hierarchies of the individual masses.

The CMB is also sensitive to the properties of dark energy (see, e.g., \cite{2010MNRAS.405.2639J}). Thanks to the current generation of experiments we know with high accuracy its energy density. The challenge for the next generation is to reveal the nature of this mysterious component. For example, a crucial step to identifying the mechanism driving cosmic acceleration will be to investigate any possible deviations in the equation of state, the ratio of pressure and energy density, from the value $w=-1$ predicted by a cosmological constant. 
Dark energy affects the CMB because it alters the Universe's expansion and it consequently changes the distance to the last scattering surface. Furthermore different dark energy models lead to different rates of growth of large scale structure which are tested by CMB lensing. 
However, dark energy properties are strongly degenerate with other geometrical parameters such as $H_{0}$. For this reason CMB provides limited information about dark energy on its own. However CMB plays an important role in dark energy studies when combined with low redshift probes. As with the neutrino sector, dark energy constraints from the CMB 
will be improved by external experiments.

In the cases of both neutrinos and dark energy then, an important lingering question is as follows: What external information can be used to improve projected constraints by breaking degeneracies?
Here, we study the dependence of projected constraints from CMB on the external priors assumed. The canonical set of parameters is as follows: cold dark matter density $\Omega_ch^2$, baryon density $\Omega_bh^2$, amplitude of primordial perturbations $A_s$, slope of primordial spectrum $n_s$, optical depth to the last scattering surface $\tau$, Hubble constant $H_0$, and sum of the neutrino masses $\sum m_\nu$. Together with extensions (the number of relativistic degrees of freedom $N_{\rm eff}$ and dark energy equation of state $w$) to the neutrino and dark energy sectors, we quantify the CMB constraints as a function of external priors. This extends the work of \citet{wu:2014}, which worked with a few fixed external priors, by quantifying the extent to which external information will improve the constraining power of a CMB-S4 experiment. 

This paper is organized as follows: in \refsec{methods} we introduce the technique and assumptions used to derive the effect of external priors on the CMB parameter constraints. In \refsec{results} we will describe our results and we then conclude with a discussion of them in \refsec{conclusions}.

\section{Assumptions and methods \label{sec:methods}}

To measure quantitatively the impact of external priors on the CMB ability to constrain cosmological parameters we use a Fisher matrix formalism. In this section we quickly review the technique and then present the chosen fiducial cosmological model together with the experimental specifications.

As usual, we define the Fisher matrix elements as the curvature of the likelihood:
\begin{equation}
	\centering
		F_{ij} \equiv - \left\langle\frac{\partial^2 \log \mathcal{L}}{\partial \theta_i \partial \theta_j} \bigg|_{\boldsymbol{\theta} = \boldsymbol{\theta_0}}\right\rangle,
	\label{eqn:Fij_def}
\end{equation}
where $\theta_{i,j}$ represent the cosmological parameters and $\boldsymbol{\theta_0}$ is the set of fiducial values that, by definition, maximizes the likelihood. 

For CMB experiments the Fisher matrix is related to the angular power spectra $\boldsymbol{C}_\ell$ by:
\begin{equation}
 F_{ij} = \sum_\ell \frac{2\ell+1}{2} f_{\rm sky} {\rm Tr} \left(  \boldsymbol{C}^{-1}_\ell( \theta) \frac{\partial \boldsymbol{C}_\ell}{\partial \theta_i} \boldsymbol{C}^{-1}_\ell( \theta) \frac{\partial \boldsymbol{C}_\ell}{\partial \theta_j}  \right)
 \label{eqn:Fij_def2}
 \end{equation}
 where $f_{\rm sky}$, the fraction of sky covered, is set to 0.75 throughout\footnote{The exact specifications of CMB-S4 are to be determined; we choose this value of $f_{\rm sky}$ to afford easy comparison with the results of \cite{wu:2014}.}
In this work we constrain cosmological parameters with CMB temperature and E-mode polarization together with the reconstructed lensing potential of large scale structure. Therefore, $\boldsymbol{C}_\ell$ in \refeq{Fij_def2} is:
 \begin{eqnarray}
 	\centering
		\mathbf{C}_\ell \equiv \left( \begin{array}{ccc}C_\ell^{TT} + N_\ell^{TT} & C_\ell^{TE} & C_\ell^{T\phi} \\ C_\ell^{TE} & C_\ell^{EE} + N_\ell^{EE} & 0 \\ C_\ell^{T\phi} & 0 & C_\ell^{\phi} + N_\ell^{\phi}\end{array}\right).
	\label{eqn:cov_definition}
\end{eqnarray}
The terms $N_\ell^{X}$ represent the instrumental noise power of the specific experiment and will be described at the end of this section.
Note that we are neglecting the term $C_\ell^{E\phi}$ since it contains very little information while adding possible numerical issues~\cite{wu:2014,2013PhRvD..87h3008H}.
Furthermore, as in \cite{wu:2014}, we use \textit{unlensed} spectra and Gaussian covariances in \refeq{Fij_def2}. 
This means that we neglect eventual imperfect reconstruction of the unlensed spectra (delensing) and the non-Gaussian effects described in \cite{benoit-levy:2012} .
However, we expect these approximations to have a small impact on the final results and, for this reason, a more careful modeling is beyond the scope of this work. For example, we checked that using lensed spectra will change the final constraint by less than 10$\%$.

The projected error on the parameter $\theta_i$, marginalized over all the other parameters, $\sigma_i$, is then:
\begin{equation}
\sigma_i \equiv \sigma (\theta_i) \geq \sqrt{(\mathbf{ F^{-1}})_{ii}}.
\label{eqn:cramer-rao}
\end{equation}
We can introduce external priors on cosmological parameters.
Indeed we simply need to add to the Fisher matrix elements the external priors (before we perform the matrix inversion of \refeq{cramer-rao}).
For example, a $1\%$ prior on the parameter $i$ can be added by:
\begin{equation}
F_{ii} \rightarrow F_{ii} + \frac{1}{(1\% \times  \theta_{i,\text{fid}})^2}.
\end{equation}

We compute the power spectra $C_{\ell}$ in \refeq{Fij_def2} using CAMB \cite{Howlett:2012mh,Lewis:1999bs} and
the derivatives in \refeq{Fij_def2} using a 5 point finite difference formula:
this high order approach allows us to use larger step-sizes around the fiducial parameters to compute derivatives. As a consequence the differences of power spectra corresponding to different values of the parameters are big enough to ensure numerical accuracy. We also test the robustness of this calculation by varying the derivative steps in the range $2-7\%$ . The constraints change by at most $10\%$ that should then be considered as a conservative estimate of numerical uncertainties. 
Our philosophy is to consider as few extensions as possible. Given the current success of the standard models of particle physics and cosmology, one of the primary goals of CMB-S4 will be to find cracks in these models. As such, the key question is as follows: What information is needed to reliably conclude that an additional parameter is required? In our baseline model we decide to include massive neutrinos, which (i) are a small extension to the Standard Model and (ii) are known to exist. So our fiducial cosmology is flat $\nu \Lambda$CDM, with assumed 
parameters from Table 2 of the \textit{Planck} best fit \cite{planck-collaboration:2014g}, i.e. $\Omega_c h^2 = 0.12029$, $\Omega_b h^2 = 0.022068$, $A_s = 2.215\times10^{-9}$ at $k_0 = 0.05\ {\rm Mpc}^{-1}$, $n_s = 0.9624$, $\tau = 0.0925$, $H_0 = 67.11$ km/s/Mpc, supplemented by an choice of $\sum m_\nu$ $\simeq$ 85\ meV to be consistent with \cite{wu:2014}.  
We then extend the parameter space by introducing $N_{\rm eff}$ and $w$ as free parameters, in each case keeping the other parameter fixed.
The fiducial values of these are $N_{\rm eff}=3.046$ and a cosmological constant equation of state, $w=-1$. In this same spirit, we keep the curvature fixed to zero, as the level predicted in the vast majority of inflationary models is of order $10^{-5}$, much too small to have an impact on the parameters we consider. When we vary $N_{\rm eff}$ we adjust the helium abundance to its canonical value from Big Bang Nucleosynthesis, which also depends on the baryon density.

The instrumental noise power $N_{\ell}^{T,E}$ and the lensing reconstruction $N_{\ell}^{\phi}$ in \refeq{Fij_def2} correspond to the optimistic level for CMB-S4 assumed in ~\cite{2013arXiv1309.5383A,wu:2014,2013PhRvD..87h3008H}.
For the temperature and E-mode polarization of the CMB, together with the improved depth and resolution we also assume that large scale foregrounds, like dust, are under control or negligible. This allows us to use all the polarization power spectrum multipoles out to $\ell_{\rm E,max}=4500$. We deal with the Poisson noise from point sources in the temperature signal by simply discarding all small scale modes with $\ell>\ell_{\rm T,max}=3000$.
The remaining source of noise, the instrumental noise, is added to the power spectrum in the usual way:
 \begin{equation}
 	\centering
		N^{X}_\ell = s^{\, 2} \exp \left(\ell(\ell+1) \frac{\theta^{\ 2}_{\textsc{fwhm}}}{8\log2}\right),
	\label{eq:beamnoise}
\end{equation}
where $\theta^{\ 2}_{\textsc{fwhm}}$ is the so called full width half maximum (FWHM) of the experiment's beam and $s$ represents the instrumental white noise.
We use a level of noise $s = 0.58$ $\mu$K-arcmin for $X=T$ and a beam of $\theta_{\textsc{fwhm}}=1$ arcmin. This corresponds to the $N_{\rm det}=10^{6}$ case of \cite{wu:2014}. 
Note that the quoted noise refers to temperature and we assume that $s \rightarrow s\times \sqrt{2}$ in the case of polarization $ XX' = \{ EE, BB \}$. 
The noise $N_\ell^{\phi\phi}$ associated with the reconstructed $\phi$ spectrum is modeled assuming an iterative reconstruction technique \cite{seljak:2004}. 

We also include information from current Baryon Acoustic Oscillation experiments and projected constraints from low-$l$ Planck polarization, as this is or will shortly be available.
For current BAO data (labeled BAO15), we follow \cite{allison:2015} and include 6dFGRS \cite{beutler:2011}, SDSS MGS \cite{ross:2015} together with LOW-Z and C-MASS \cite{anderson:2014}.
The noise assumed for Planck corresponds to the ``Planck-pol'' specifications of \cite{allison:2015}, where noise levels were approximated by scaling the current sensitivities according to the Planck Blue Book.

\section{Results \label{sec:results}}
In this section we present our main results. Starting from $\nu \Lambda$CDM we then explore extensions to this model by considering constraints on $N_{\rm eff}$ and $w$.

\subsection{Neutrino Masses, $\sum m_\nu$}

Small scale structure formation is suppressed if fast-moving neutrinos comprise a significant part of the matter budget. 
Indeed, below the free-streaming scale, the matter power spectrum is suppressed in the presence of
massive neutrinos by a factor $\Delta P/P\simeq -8f_{\nu}$ , where $f_{\nu} = \Omega_{\nu} / \Omega_{m}$ is the contribution of neutrinos to the total matter density.
This effect is probed by CMB lensing and, considering the robust prediction from the particle physics Standard Model for the number density of active neutrinos, it can be directly transformed into a constraint on the sum of the neutrino masses $\sum m_\nu$. This is particularly exciting because we now know that neutrinos are massive, with a lower limit $\sum m_\nu>50$ meV that emerges from oscillation experiments (see \cite{bellini:2013} and references therein).

%==========
\begin{figure}[htbp]
\begin{center}
\includegraphics{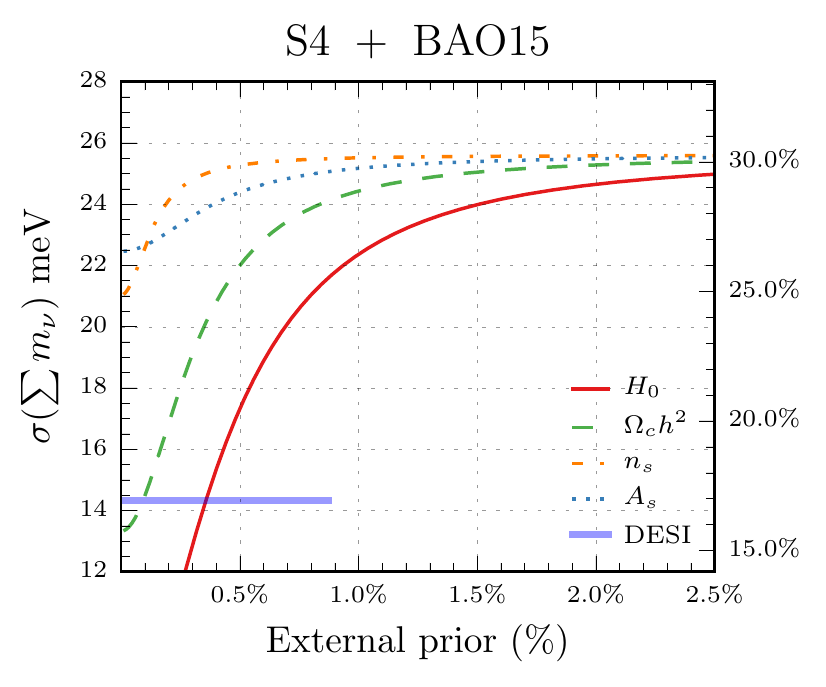}
\caption{Projected constraints on the sum of the neutrino masses (absolute on the left and relative on the right) from CMB-S4 and current BAO measurements as a function of priors on the other cosmological parameters. For comparison, the {\it internal} (marginalized constraints from CMB-S4 and current BAO) constraints on each parameter are:  
$\sigma_{\rm pipeline}(H_{0})/H_{0}=0.6\%$, 
$\sigma_{\rm pipeline}(\Omega_{c}h^{2})/\Omega_{c}h^{2}=0.4\%$,
$\sigma_{\rm pipeline}(n_{s})/n_{s}=0.1\%$,
$\sigma_{\rm pipeline}(A_{s})/A_{s}=0.4\%$. 
The horizontal (blue) line corresponds to CMB S4 experiment + Planck Pol + DESI.}
\label{fig:prior_omeganuh2-h}
\end{center}
\end{figure}
%==========
%==========

\begin{figure}[htbp]
\begin{center}
\includegraphics{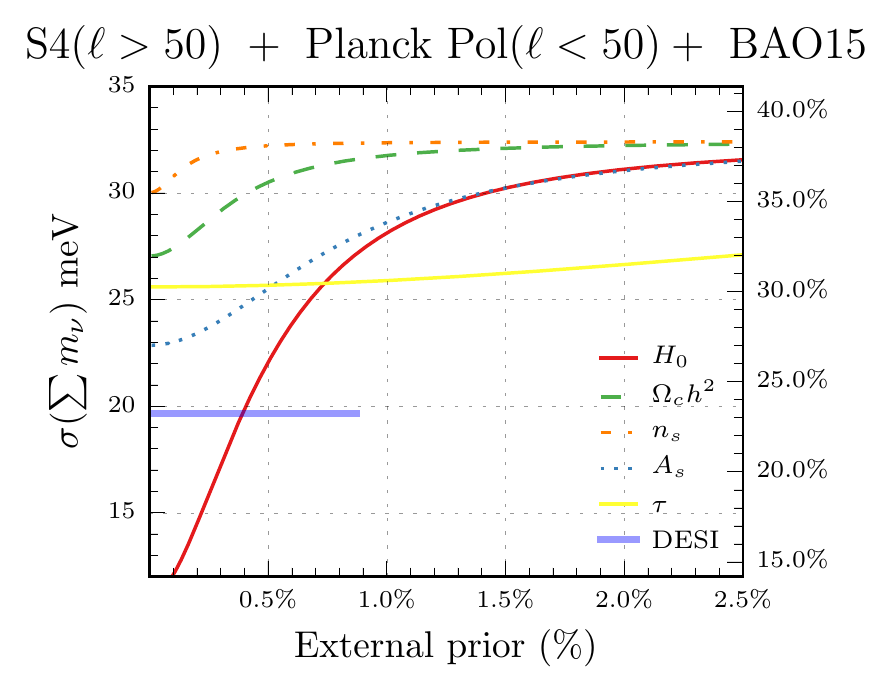}
\caption{Same as Fig.~1, but with $l<50$ modes excluded from CMB-S4, instead coming from Planck. Note the importance of obtaining information on the optical depth $\tau$ if the low-$l$ modes are not measured accurately.
For comparison, the {\it internal} (marginalized constraints from CMB-S4 and current BAO) constraints on each parameter are:  
$\sigma_{\rm pipeline}(H_{0})/H_{0}=0.6\%$, 
$\sigma_{\rm pipeline}(\Omega_{c}h^{2})/\Omega_{c}h^{2}=0.4\%$,
$\sigma_{\rm pipeline}(n_{s})/n_{s}=0.1\%$,
$\sigma_{\rm pipeline}(A_{s})/A_{s}=0.9\%$,
$\sigma_{\rm pipeline}(\tau)/\tau=5\%$.}
\label{fig:prior-50}
\end{center}
\end{figure}
%==========

%==========

\reffig{prior_omeganuh2-h} shows the projected errors on $\sum m_\nu$ from CMB-S4 (+Planck+current BAO) as a function of external priors on $H_{0}, \Omega_{c}h^{2}, n_s$ and $A_s$. 
Without any external priors (far right in figure), upcoming CMB experiments are poised to obtain 1-sigma limits on $\sum m_\nu$ of order 25 meV, corresponding to a 2-sigma detection even in the worst case scenario. Adding in external priors helps this significantly though, as discussed, for example, in \cite{di-valentino:2015,2013arXiv1309.5383A,pan:2015a,allison:2015}. 
There, the added prior was ``DESI BAO'', representing a measurement of distances as a function of redshift from the baryon acoustic oscillation feature probed by the planned experiment DESI \cite{levi:2013}. This is represented by the light blue band in our figures. In addition to showing the implications of introducing this fixed external constraint, \reffig{prior_omeganuh2-h} generalizes the extent to which external information will improve the neutrino mass constraints. For example, in \reffig{prior_omeganuh2-h} the red solid curve shows the improvement as additional information on the Hubble constant is applied. We can see, for example, that imposing {\it external} information at the level of $\sigma_{H_0}/H_0=0.5\%$ would improve the neutrino mass constraint to 17 meV. It is not surprising that this is close to the constraint obtained with DESI because the BAO distances in this model depend most sensitively on $\Omega_ch^2$ and $H_{0}$.
Similarly, adding to CMB S4 + BAO15 an external constraint on the dark matter density equal to its internal constraint of $0.4\%$ improves the projected neutrino mass constraint to 20 meV. These are multi-dimensional examples of the familiar fact that adding two sets of equal and independent constraints (in this case internal to S4+BAO15 and external) reduces the overall error by a factor of $\sqrt{2}$. 

The physical reason underlying the improvement from $\Omega_ch^2$ is simple: lensing constrains $f_\nu$, the ratio of the neutrino density to the matter density. In order to extract the neutrino density (and therefore the neutrino mass), we need to know the matter density. A prior on $H_{0}$ helps for a related, but subtle reason. 
The position of the CMB peaks, well constrained by measurement, in a flat Universe depends on the combination $\Omega_ch^{2.93}$ \cite{planck-collaboration:2014}.
Therefore, an external prior on $H_{0}$ will improve the internal reconstruction of $\Omega_ch^2$, leading back to tighter neutrino mass constraints.
As found in \cite{2013arXiv1309.5383A,pan:2015a,allison:2015}, the projected error falls below 20 meV with these priors. 
The left-hand side of \reffig{prior_omeganuh2-h} shows that even tighter external constraints on either $H_0$ or $\Omega_ch^2$ would significantly improve the neutrino mass constraints.

%==========

Because of the presence of foregrounds, it may be difficult to get the very low-$\ell$ modes from the ground with CMB-S4, so in \reffig{prior-50}, we limit the CMB S4 modes to $\ell>\ell_{\rm min}=50$. For the large scale modes $\ell<\ell_{\rm min}=50$ we use Planck at the level of accuracy forecasted by the Planck Blue Book.  
Together with the importance of $H_{0}$ and $\Omega_ch^2$ previously observed, in this case external priors on $A_{s}$ and $\tau$ are found to be important. 
Indeed, as recently pointed out by \cite{allison:2015}, a degeneracy exists between $\sum m_\nu$ and $A_{s}$ and, as a consequence, between $\sum m_\nu$ and $\tau$. This can be understood as follows.
The CMB lensing reconstruction is noisy at scales larger than the neutrino free streaming scale (see Fig. 5 of \cite{2013arXiv1309.5383A}). For this reason it is hard to measure the unsuppressed CMB lensing power spectra that is then compared to the small scale suppressed one to constrain the values of $f_{\nu}$. The unsuppressed amplitude of the lensing spectrum is actually better constrained by the primordial amplitude $A_{s}$, which is probed by the primordial CMB. Unfortunately, the CMB spectra are sensitive to $A_{s}\exp^{-2\tau}$, so a measurement of the optical depth is required to infer $A_{s}$ and therefore to tighten the constraints on $\sum m_\nu$. If CMB S4 will not be designed to measure scales $\ell<50$ to avoid difficulties in dealing with atmospheric emission, an external prior on the optical depth will be crucial. This can be obtained from future 21 cm surveys \cite{liu:2015} or other CMB experiments, satellites like LiteBIRD \cite{matsumura:2014}, PIXIE\cite{kogut:2011}, and COrE/PRISM \cite{the-core-collaboration:2011,prism-collaboration:2013}, balloon-borne experiments like SPIDER \cite{filippini:2010}, EBEX \cite{reichborn-kjennerud:2010} or eventually ground experiments like CLASS \cite{essinger-hileman:2014}.

\subsection{Relativistic Degrees of Freedom, $N_{\rm eff}$}

In the standard model of cosmology, three active neutrinos are thermally produced in the early Universe. Were they to decouple well before the epoch of electron-positron annihilation, their energy density after their decoupling would be equal to $3\times (7/8)\times (4/11)^{4/3}\,\rho_{\rm cmb}$, with the first factor capturing the contributions from the 3 active species; the second the difference between fermions and bosons, and the last the relative heating of the photons in the CMB by electron-positron annihilation. However, decoupling is not a discrete event and occurs close to the time of electron-positron annihilation, so the neutrinos share a bit in the heating, with the factor of 3 replaced by $N_{\rm eff}=3.046$. The additional fraction depends not only on well-known neutrino scattering rates but also on finite temperature quantum corrections. 
Upcoming experiments have the potential to measure this tiny deviation of $N_{\rm eff}$ from 3. This will be a precision test of our understanding of the Universe when it was about a second old. It also will constrain Beyond the Standard Model physics. For example, as pointed out in \cite{baumann:2015}, tight constraints on $N_{\rm eff}$ will allow us to rule out new particles with couplings that enabled them to thermalize early on but that decoupled when the temperature was above $\sim$100 GeV.

\begin{figure}[htbp]
\begin{center}
\includegraphics{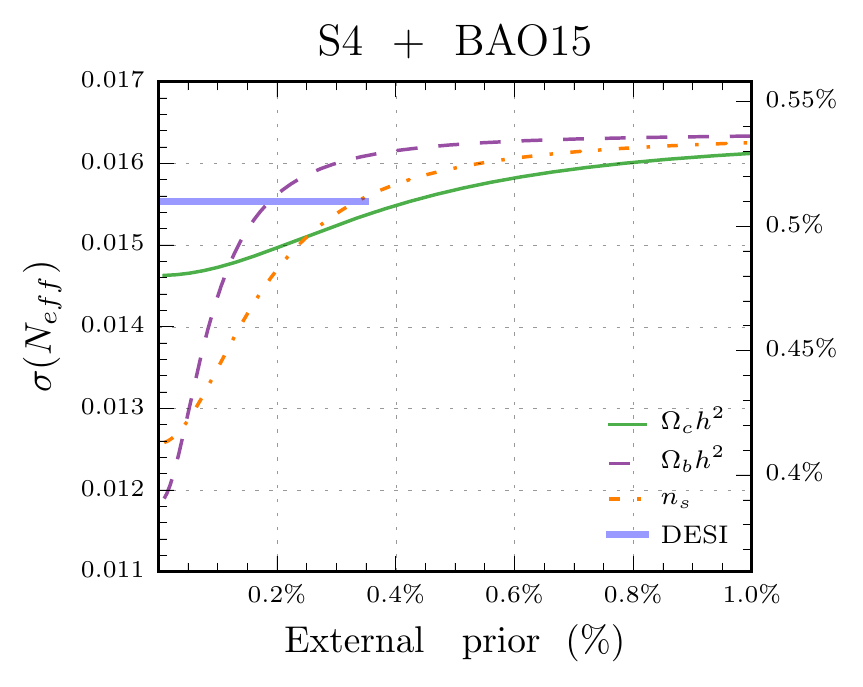}
\caption{Constraints on $N_{\rm{eff}}$ (absolute on the left and relative to its value on the right) as a function of priors on $\Omega _{b} h^{2}$, $\Omega _{c} h^{2}$ and $n_{s}$. 
For comparison, the {\it internal} (marginalized constraints from CMB-S4 and current BAO) constraints on each parameter are:  
$\sigma_{\rm pipeline}(\Omega_{c}h^{2})/\Omega_{c}h^{2}=0.4\%$,
$\sigma_{\rm pipeline}(n_{s})/n_{s}=0.1\%$,
$\sigma_{\rm pipeline}(\Omega_{b}h^{b})/\Omega_{c}h^{2}=0.1\%$,
The horizontal (blue) line corresponds to CMB S4 experiment + Planck Pol + DESI.}
\label{fig:prior_neff}
\end{center}
\end{figure}

\reffig{prior_neff} shows projections for how well CMB-S4 will do at measuring $N_{\rm eff}$ as a function of priors on $H_{0}$, $\Omega_{b}h^{2}$ and $\Omega_{c}h^{2}$ (the sum of the neutrino masses must be included as a free parameter). With no priors on the other parameters, the projected $1$-sigma error is $0.016$, close to a 3-sigma detection of the expected deviation from $N_{\rm eff}=3$. 

The sensitivity to relativistic degrees of freedom comes from the effect of extra species on the damping tail of the CMB anisotropies \cite{2013PhRvD..87h3008H}, both in temperature and polarization, so parameters that also strongly affect this part of the spectrum, like the slope $n_s$ and the baryon density $\Omega_bh^2$, are the most relevant to improve the $N_{\rm eff}$ constraint. As such, the blue and green dashed lines in \reffig{prior_neff} show that obtaining external priors on either of these would reduce the errors on $N_{\rm eff}$ to get close to a 4-sigma detection of the partial decoupling prediction. Besides futuristic high redshift 21 cm ideas, the best bet to improve the constraints on $n_s$ are probes that push to redshift $z>1$ such as the Lyman alpha forest. There are many more linear Fourier modes available at high $z$ that can be used to obtain the larger level arm from large to small scales, which is essential to extract high precision measurements of $n_s$. 

\subsection{Dark Energy Equation of State, $w$}

The CMB constrains the late-time dark energy equation of state in two ways. The observed CMB spectra are very sensitive to the distance to the last scattering surface which depends on $w$. 
Furthermore CMB lensing probes the growth of structure at late times which is different for different dark energy equations of state.

\begin{figure}[htbp]
\begin{center}
\includegraphics{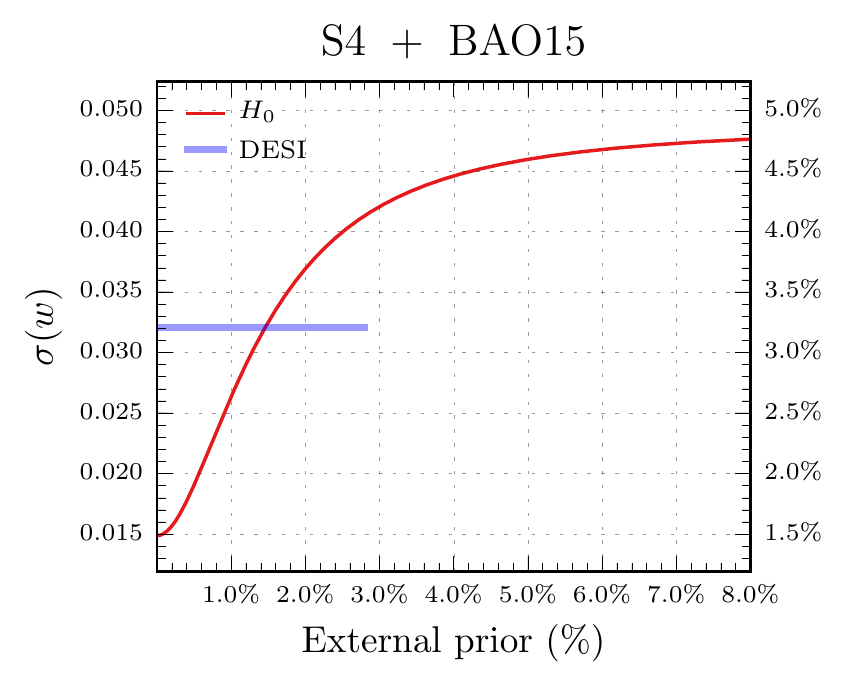}
\caption{Constraints on $w$ (absolute on the left and relative to its value on the right) as a function of priors on $H_{0}$. 
For comparison, the {\it internal} (marginalized constraints from CMB-S4 and current BAO) constraints on $H_{0}$ when w is free to vary is 
$\sigma_{\rm pipeline}(H_{0})/H_{0}=1.9\%$.
The horizontal (blue) line corresponds to CMB S4 experiment + Planck Pol + DESI.}
\label{fig:prior_w-h}
\end{center}
\end{figure}

%==========

\begin{figure}[htbp]
\begin{center}
\includegraphics{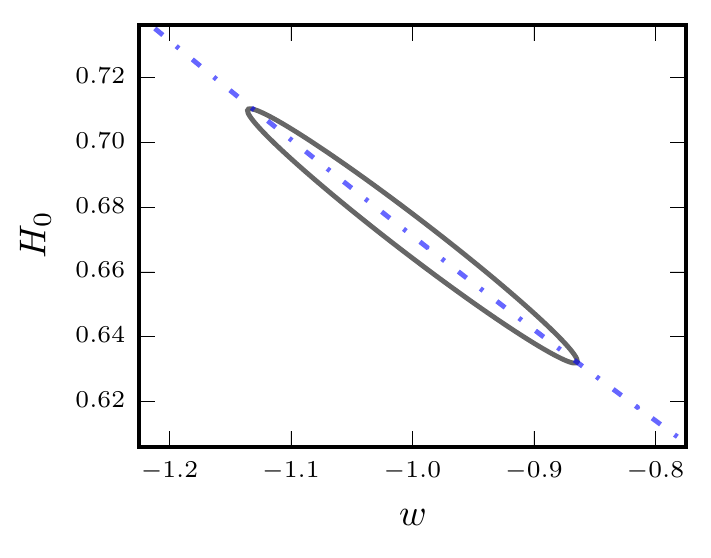}
\caption{Two dimensional constraint (CMB S4 + Planck) 1$\sigma$ ellipse for $w,H_{0}$ space. The dashed (blue) line corresponds to values in the plane corresponding to the same distance to the last scattering surface.} 
\label{fig:ellipse_w_H0}
\end{center}
\end{figure}
%==========

\reffig{prior_w-h} shows that without any priors, CMB experiments (+BAO15) will do better than current constraints, which hover around 10\%. However, an external prior on $H_{0}$ two times more accurate than CMB S4 (i.e. $1\%$) can improve the CMB constraint on $w$ by a factor of two. 
This is primarily due to the fact that CMB constrains $w$ through a precise measurement of the comoving distance to the last scattering surface. However that can be kept fixed while varying $w$ by accordingly changing the value of $H_{0}$ (and $\Omega_{\Lambda}$ to keep the Universe flat). Indeed in a two dimensional ($H_{0},w$) space the CMB constraint approximately lies on the region of constant large scattering surface distance as shown in \reffig{ellipse_w_H0}.

\section{Conclusions \label{sec:conclusions}}
A CMB-S4 experiment will measure the sum of the neutrino masses, constrain extra relativistic degrees of freedom predicted in many extensions of the Standard Model, and measure the dark energy equation of state at the few percent level. 
These results can be improved even further by combining them with information from external experiments.
Understanding the impact of such external information is crucial in determining the experimental specifications and requirements needed for CMB-S4, and ultimately in maximizing the scientific impact of the future CMB experiments. 

Toward that goal, we investigated here how these constraints depend on external priors:
\begin{description}
\item[$\sum m_\nu$] To improve the CMB constraint on the sum of neutrino masses by a factor of two, low-redshift measurements will be needed. Indeed an external prior on $\Omega_{c}h^{2}$ and $H_{0}$ at the level attained by CMB-S4 (sub-percent) can reduce the 1-$\sigma$ uncertainty by almost a factor of two.
Furthermore, if the final design of CMB S4 does not include measurements on large scales ($\ell<50$), external constraints on the optical depth $\tau$, from 21 cm experiments or CMB experiments measuring the large
scales, will become very important.
\item[$N_{\rm eff}$] The constraint on the relativistic degrees of freedom will benefit from external priors on parameters that affect the shape of the damping tail, such as the spectral index $n_s$. These are most likely to come from surveys that measure a large fraction of the $z>1$ linear Fourier modes.
\item[$w$] In this case the improvement will mainly come from external priors on $H_{0}$. For example a $1\%$ prior from an external experiment will reduce the error on $w$ by a factor of two.\end{description}

Much of the improvement that would come on neutrino masses and $w$ from $H_0$ or $\Omega_ch^2$ priors is anticipated to be provided by distance measurements made by DESI. But, additional $H_0$ measurements beyond DESI would help, as evidenced by the sharply falling curves in the left-most regions of \reffig{prior_omeganuh2-h}, \reffig{prior-50}, and \reffig{prior_w-h}. Furthermore, perhaps the most tantalizing finding is that future surveys beyond DESI and LSST \cite{lsst-dark-energy-science-collaboration:2012} that would tighten the constraints on, say, $n_s$ could indirectly lead to constraints in $N_{\rm eff}$ much less than a percent. Given the power of these constraints to probe physics beyond the Standard Model, such surveys seem particularly interesting to pursue.

\begin{acknowledgments}
We thank Wayne Hu for useful discussions. AM wants to thank Zhen Pan who allowed a careful cross-check of our results.
This work was partially supported by the Kavli Institute for Cosmological Physics at the University of Chicago through grants NSF PHY-1125897 and an endowment from the Kavli Foundation and its founder Fred Kavli.
%%%=================================================================
The work of SD is supported by the U.S. Department of Energy, including grant DE-FG02-95ER40896.
\end{acknowledgments}

\bibliography{N_eff_prior_paper}

\end{document}